%% file: paper.tex
\documentclass[letterpaper, 10 pt, conference]{ieeeconf}

\IEEEoverridecommandlockouts
\usepackage{cite}
\usepackage{amsmath,amssymb,amsfonts}
\usepackage{graphicx}
\usepackage{textcomp}
\usepackage{xcolor}
\usepackage[utf8]{inputenc}
\usepackage[T1]{fontenc}
\usepackage{mathtools}
\DeclarePairedDelimiter{\norm}{\lVert}{\rVert}
\usepackage{bm}
\usepackage{float}
\usepackage{mathrsfs}
\usepackage{relsize}
\usepackage{hyperref}
\usepackage{amsthm}
\newenvironment{compactproof}{%
  \begin{proof}\begingroup
    \setlength{\abovedisplayskip}{2pt}
    \setlength{\belowdisplayskip}{2pt}
    \setlength{\abovedisplayshortskip}{2pt}
    \setlength{\belowdisplayshortskip}{2pt}
    \linespread{0.85}\selectfont
}{%
    \endgroup\end{proof}
}

\newtheorem{theorem}{Theorem}
\newtheorem{lemma}{Lemma}
\newtheorem{proposition}{Proposition}
\newtheorem{corollary}{Corollary}
\newtheorem{definition}{Definition}
\theoremstyle{remark}
\newtheorem{remark}{Remark}

\DeclareMathAlphabet{\mathcal}{OMS}{cmsy}{m}{n}

\graphicspath{{images/}}

\begin{document}

\title{Data-Driven Reachability Analysis with Optimal Input Design}

\author{Peng Xie, Davide M.~Raimondo, Rolf Findeisen, Amr Alanwar%
\thanks{P.~Xie and A.~Alanwar are with the Department of Computer Engineering,
TUM School of Computation, Information and Technology,
Technical University of Munich, 74076 Heilbronn, Germany.
\texttt{(e-mail: p.xie@tum.de, alanwar@tum.de)}}%
\thanks{Rolf Findeisen is with the Technical University of Darmstadt, 64283 Darmstadt, Germany.
\texttt{(e-mail: rolf.findeisen@iat.tu-darmstadt.de)}}
\thanks{D.~M.~Raimondo is with the Department of Engineering and Architecture,
University of Trieste, 34127 Trieste, Italy.
\texttt{(e-mail: davidemartino.raimondo@dia.units.it)}}%
}

\maketitle

\begin{abstract}
This paper addresses data-driven reachability analysis for discrete-time
linear systems subject to bounded process noise, where the
system matrices are unknown and only input--state trajectory data are
available. Building on the constrained matrix zonotope (CMZ) framework, two complementary strategies are proposed to reduce conservatism in reachable-set over-approximations. First, the standard Moore--Penrose pseudoinverse is replaced with a row-norm-minimizing right inverse computed via a second-order cone program, yielding tighter generators and less conservative reachable sets. Second, an online A-optimal input design strategy is introduced to improve the informativeness of the collected data and to reduce the uncertainty of the resulting model set. The proposed framework extends naturally to piecewise affine systems through mode-dependent data partitioning.
Numerical results on a five-dimensional stable LTI system and a two-dimensional piecewise affine system demonstrate that combining designed inputs with the row-norm right inverse significantly reduces conservatism compared to a baseline using random inputs and the pseudoinverse.

\end{abstract}

\input{sections/introduction}

\input{sections/preliminaries}

\input{sections/method}

\input{sections/theory}

\input{sections/experiments}

\input{sections/conclusion}

\bibliographystyle{ieeetr}

\input{paper.bbl}
\end{document}

%% file: sections/introduction.tex
\section{Introduction}\label{sec:intro}

Safety verification of dynamical systems requires computing all states reachable from prescribed initial conditions under all admissible inputs and disturbances. When the system dynamics are known and disturbances are bounded, classical reachability analysis tools propagate set-valued representations---such as zonotopes, polytopes, or ellipsoids---forward in time to obtain guaranteed over-approximations of the true reachable set~\cite{girard2005reachability,althoff2010reachability,kuhn1998rigorously,%
althoff2010computing}.
In many practical applications, however, the system model is unavailable or difficult to obtain, motivating data-driven approaches that bypass explicit system identification and instead construct set-valued models directly from measured trajectories~\cite{willems2005note,depersis2020formulas,waarde2020data}.

Alanwar et al.~\cite{amr23reachable} introduced a matrix-zonotope framework
for data-driven reachability analysis of discrete-time linear systems with bounded
noise.  In this framework, the set of system matrices consistent with
the observed data and the noise bounds is represented as a matrix zonotope,
and reachable sets are propagated by multiplying this model set with the current state set while accounting for bounded noise.  A right inverse of the data matrix is used to construct the consistent model set, with the Moore--Penrose pseudoinverse as the default choice. The resulting over-approximation is sound but can be conservative, particularly when the collected data are uninformative.

This conservatism can be reduced along three directions. First, standard matrix zonotopes do not exploit the kernel of the data regressor, which induces equality constraints that can tighten the model set. Second, the choice of right inverse influences the size of the resulting model set and can be optimized to reduce conservatism.
Third, active input design~\cite{hjalmarsson2005experiment,scott2014input,marseglia2017active} can improve the informativeness of the collected data, leading to better-conditioned right inverses and smaller model sets.

The first direction has been addressed in~\cite{alanwar2022data} through the constrained matrix zonotope (CMZ) framework, which enforces kernel-consistency constraints derived from the nullspace of the data regressor and yields a tighter outer approximation of the model set. Building on this framework, the present paper addresses the remaining two directions, namely right-inverse optimization and active input design. A row-norm-minimizing right inverse is introduced, computed via second-order cone programming (SOCP), which yields smaller generators compared to the pseudoinverse (Theorem~\ref{prop:row_norm}). An online A-optimal input design strategy over constrained zonotope input sets is further proposed, combining uniform sampling for exploration with SQP-based refinement. The proposed approach extends naturally to piecewise affine (PWA) systems through mode-dependent data partitioning and hybrid zonotope propagation~\cite{bird2023hybrid}.

The remainder of the paper is organized as follows. Section~\ref{sec:prelim} introduces notation and set-theoretic definitions. Section~\ref{sec:method} formalizes the problem and presents the proposed approach, including the matrix-zonotope model-set construction, the constrained matrix zonotope tightening, the row-norm right inverse, and the A-optimal input design. Section~\ref{sec:theory} presents the main theoretical results. Section~\ref{sec:experiments} reports numerical experiments. Section~\ref{sec:conclusion} concludes the paper.

%% file: sections/preliminaries.tex
\section{Preliminaries and Definitions}\label{sec:prelim}

Matrices are denoted by uppercase letters ($A$, $B$), vectors by lowercase letters ($x$, $c$), and sets by calligraphic letters ($\mathcal{Z}$,
$\mathcal{W}$).  The identity matrix is $I_d \in \mathbb{R}^{d \times d}$.
The pseudoinverse of $M$ is $M^\dagger$, and
$\mathrm{rank}(M)$ denotes its rank.
The Frobenius, Euclidean, and infinity norms are denoted by
$\norm{\cdot}_F$, $\norm{\cdot}_2$, and  $\norm{\cdot}_\infty$ respectively.  The operator $\mathrm{vec}(\cdot)$ stacks the columns of a matrix into a vector.
The unit infinity-norm ball in $\mathbb{R}^n$ is
$\mathcal{B}_\infty^n := \{\xi \in \mathbb{R}^n \mid \norm{\xi}_\infty \le 1\}$.
The canonical basis vector $e_t \in \mathbb{R}^T$ has a one in position~$t$
and zeros elsewhere.
The Minkowski sum of two sets is
$\mathcal{A} \oplus \mathcal{B} := \{a + b \mid a \in \mathcal{A},\;
b \in \mathcal{B}\}$,
and the Cartesian product is
$\mathcal{A} \times \mathcal{B} := \{(a,b) \mid a \in \mathcal{A},\;
b \in \mathcal{B}\}$.
The determinant and trace of a square matrix are
$\det(\cdot)$ and $\mathrm{tr}(\cdot)$, respectively.
The minimum singular value of~$M$ is $\sigma_{\min}(M)$.

\begin{definition}[Zonotope~\cite{girard2005reachability}]\label{def:zono}
A zonotope $\mathcal{Z} \subset \mathbb{R}^n$ with center
$c \in \mathbb{R}^n$ and generator matrix $G \in \mathbb{R}^{n \times m}$
is defined as
$\mathcal{Z} = \langle c, G \rangle
:= \{c + G\xi \mid \norm{\xi}_\infty \le 1\}$.
\end{definition}

\begin{definition}[Constrained Zonotope~\cite{scott2016constrained}]\label{def:conzono}
A constrained zonotope $\mathcal{C} \subset \mathbb{R}^n$ with
center $c$, generator matrix $G$, and  linear constraints defined by
$A_c \in \mathbb{R}^{p \times m}$ and 
$b_c \in \mathbb{R}^p$ is
\begin{equation}\label{eq:conzono_def}
\mathcal{C} = \langle c, G, A_c, b_c \rangle
:= \{c + G\xi \mid \norm{\xi}_\infty \le 1,\; A_c\xi = b_c\}.
\end{equation}
Zonotopes are a special case of Constrained Zonotopes (CZs) without equality constraints.
CZs are closed under Minkowski sum, linear maps, and
Cartesian products~\cite{scott2016constrained}.
\end{definition}

\begin{definition}[Matrix Zonotope~\cite{amr23reachable}]\label{def:matzono}
A matrix zonotope $\mathcal{M} \subset \mathbb{R}^{n \times m}$ with
center $C \in \mathbb{R}^{n \times m}$ and generator matrices
$G_\ell \in \mathbb{R}^{n \times m}$, $\ell = 1,\dots,\kappa$,
 is the set
\begin{equation}\label{eq:matzono_def}
\mathcal{M} = \langle C, \{G_\ell\}_{\ell=1}^\kappa \rangle
:= \Big\{C + \sum_{\ell=1}^\kappa \beta_\ell G_\ell
\;\Big|\; \beta \in [-1,1]^\kappa\Big\}.
\end{equation}
\end{definition}

\begin{definition}[Constrained Matrix Zonotope~\cite{alanwar2022data}]\label{def:conmatzono}
A constrained matrix zonotope
$\mathcal{M}_c \subset \mathbb{R}^{n \times m}$ augments a matrix
zonotope with linear equality constraints on the coefficient vector:
\begin{multline}\label{eq:conmatzono_def}
\mathcal{M}_c = \langle C, \{G_\ell\}_{\ell=1}^\kappa,
A_{\mathrm{cmz}}, b_{\mathrm{cmz}} \rangle
:= \Big\{C + \textstyle\sum_{\ell=1}^\kappa \beta_\ell G_\ell
\;\Big|\\
\beta \in [-1,1]^\kappa,\;
A_{\mathrm{cmz}}\beta = b_{\mathrm{cmz}}\Big\}.
\end{multline}
Every matrix zonotope is a constrained matrix zonotope with
$A_{\mathrm{cmz}} = []$.  Because the constraints restrict the feasible
$\beta$, it follows that $\mathcal{M}_c \subseteq \mathcal{M}$.
\end{definition}

\begin{proposition}[CMZ--Zonotope Product Over-Approximation~\cite{alanwar2022data}]\label{prop:cmz_product}
Let $\mathcal{M}_c = \langle C, \{G_\ell\}_{\ell=1}^\kappa,
A_{\mathrm{cmz}}, b_{\mathrm{cmz}} \rangle$ be a constrained matrix
zonotope with $A_{\mathrm{cmz}} \in \mathbb{R}^{q \times \kappa}$, and
let $\mathcal{Z} = \langle c_z, [g_{z,1},\dots,g_{z,p}] \rangle$ be a
zonotope.  Then
\begin{multline}\label{eq:cmz_zono_mult}
\mathcal{M}_c\,\mathcal{Z}
\subseteq \Big\langle C\,c_z,\;
\{C\,g_{z,i}\}_{i=1}^{p}
\cup \{G_\ell\,c_z\}_{\ell=1}^{\kappa}\\
\cup\;\{G_\ell\,g_{z,i}\}_{\ell,i},\;
\hat{A}_{\mathrm{cmz}},\;b_{\mathrm{cmz}}\Big\rangle,
\end{multline}
where the constraint matrix
$\hat{A}_{\mathrm{cmz}}
= \big[\,0_{q \times p}\;\;\; A_{\mathrm{cmz}}\;\;\;
0_{q \times \kappa p}\,\big]$
selects the CMZ coefficients~$\beta$ from the combined coefficient
vector $(\alpha, \beta, \gamma) \in \mathbb{R}^{p+\kappa+\kappa p}$.
The result is a constrained zonotope that preserves the equality
constraints of the CMZ while treating the bilinear products
$\beta_\ell \alpha_i$ as independent factors~$\gamma_{\ell,i}$.
\end{proposition}

\begin{definition}[Hybrid Zonotope~\cite{bird2023hybrid}]\label{def:hybzono}
A hybrid zonotope $\mathcal{Z}_h \subset \mathbb{R}^n$ with center
$c_z \in \mathbb{R}^n$, continuous generators
$G_z^c \in \mathbb{R}^{n \times n_g}$, binary generators
$G_z^b \in \mathbb{R}^{n \times n_b}$, and equality constraints
$(A_z^c, A_z^b, b_z)$ is defined as
\begin{multline}\label{eq:hybzono_def}
\mathcal{Z}_h = \langle G_z^c, G_z^b, c_z, A_z^c, A_z^b, b_z \rangle
:= \Big\{c_z + G_z^c \xi^c + G_z^b \xi^b \;\Big|\\
\norm{\xi^c}_\infty \le 1,\;
\xi^b \in \{-1,1\}^{n_b},\;
A_z^c \xi^c + A_z^b \xi^b = b_z\Big\}.
\end{multline}
Hybrid zonotopes support Minkowski sum, generalized intersection, and
halfspace intersection~\cite{bird2023hybrid}, operations that are used for
piecewise affine system propagation in Section~\ref{sec:pwa}.
\end{definition}

%% file: sections/method.tex
\section{Problem Formulation and Method}\label{sec:method}

\subsection{Problem Statement and Data Model}\label{sec:problem}

The goal of this work is to compute \emph{sound} over-approximations of
exact reachable sets for discrete-time systems whose governing model is
unknown, given only bounded-noise input--state trajectories.
The true system evolves according to
\begin{equation}\label{eq:lti_true}
x(k+1) = A_{\mathrm{tr}}\,x(k) + B_{\mathrm{tr}}\,u(k) + w(k),
\end{equation}
where $x(k) \in \mathbb{R}^{n_x}$ is the state,
$u(k) \in \mathbb{R}^{n_u}$ is the input, and the process disturbance
satisfies $w(k) \in \mathcal{W}$ for all $k$.
The pair $[A_{\mathrm{tr}}\;\; B_{\mathrm{tr}}]$ is unknown.

Given an initial set $x(0) \in \mathcal{X}_0$, an input-constraint set
$u(k) \in \mathcal{U}$, and bounded process noise $w(k) \in \mathcal{W}$,
the exact reachable set at time $k$ is
\begin{multline}\label{eq:reach_exact}
\mathcal{R}_k := \big\{x(k) \mid
x(0) \in \mathcal{X}_0,\;
u(t) \in \mathcal{U},\\
w(t) \in \mathcal{W},\;
\text{\eqref{eq:lti_true} holds for }t=0{:}k{-}1\big\}.
\end{multline}
The objective is to compute data-driven over-approximations
$\widehat{\mathcal{R}}_k$ such that
$\mathcal{R}_k \subseteq \widehat{\mathcal{R}}_k$ for $k$ over a
prescribed horizon.

Instead of a model, $K$ input--state trajectories of lengths $T_i + 1$
are available:
$\{u^{(i)}(k)\}_{k=0}^{T_i-1}$ and $\{x^{(i)}(k)\}_{k=0}^{T_i}$ for
$i = 1, \dots, K$.
The shifted data matrices are
\begin{align}
X_+ &= \big[\,x^{(1)}(1) \cdots x^{(K)}(T_K)\,\big],\nonumber\\
X_- &= \big[\,x^{(1)}(0) \cdots x^{(K)}(T_K\!-\!1)\,\big],
\label{eq:data_shifted}\\
U_- &= \big[\,u^{(1)}(0) \cdots u^{(K)}(T_K\!-\!1)\,\big].\nonumber
\end{align}
The total number of one-step transitions is $T := \sum_{i=1}^K T_i$.
The input constraint set $\mathcal{U}$ is a constrained zonotope
(Definition~\ref{def:conzono}):
$\mathcal{U} = \langle c_u, G_u, A_u, b_u \rangle$,
where $c_u \in \mathbb{R}^{n_u}$, $G_u \in \mathbb{R}^{n_u \times m}$
are generator columns, and the equality constraints $(A_u, b_u)$ couple
the generator factors $\xi \in \mathbb{R}^m$.

\subsection{Data-Driven Sets of Models via Matrix Zonotopes}\label{sec:modelset}

The one-step data satisfy the matrix relation
\begin{equation}\label{eq:data_relation}
X_+ = \underbrace{[A_{\mathrm{tr}}\;\; B_{\mathrm{tr}}]}_{=: M_{\mathrm{tr}}}
\,\Phi + W_-,
\end{equation}
where $W_- := [w(0)\;\cdots\; w(T\!-\!1)] \in \mathbb{R}^{n_x \times T}$
stacks the unknown disturbances, and the regressor matrix is
\begin{equation}\label{eq:Phi_def}
\Phi := \begin{bmatrix}X_-\\ U_-\end{bmatrix}
\in \mathbb{R}^{d \times T},\qquad d := n_x + n_u.
\end{equation}
Assuming that $\Phi$ has full row rank, right inverses
$H \in \mathbb{R}^{T \times d}$ satisfying
\begin{equation}\label{eq:right_inverse}
\Phi H = I_d
\end{equation}
exist.

Let $\mathcal{M}_w$ be a matrix zonotope that over-approximates all
admissible stacked disturbances $W_-$.  The data matrix zonotope without noise
is defined as
\begin{equation}\label{eq:denoised_set}
\mathcal{N} := X_+ - \mathcal{M}_w
= \langle C_n,\, \{G_\ell\}_{\ell=1}^\kappa \rangle,
\end{equation}
with $C_n := X_+ - C_w$ and $G_\ell := -G_{w,\ell}$,
where $\mathcal{M}_w = \langle C_w, \{G_{w,\ell}\}_{\ell=1}^\kappa \rangle$.
The model-set outer approximation is then
\begin{multline}\label{eq:modelset_mz}
\mathcal{M}_\Sigma^{\mathrm{MZ}}(H) := \mathcal{N}\,H\\
= \bigg\{\bigg(C_n + \textstyle\sum_{\ell=1}^\kappa \beta_\ell G_\ell\bigg)H
\;\bigg|\; \beta \in [-1,1]^\kappa\bigg\},
\end{multline}
satisfying
$M_{\mathrm{tr}} \in \mathcal{M}_\Sigma^{\mathrm{MZ}}(H)$
under bounded-noise assumptions~\cite{amr23reachable}.

\subsection{Kernel-Consistent Constrained Matrix Zonotope Tightening}\label{sec:cmz}

Alanwar et~al.~\cite{alanwar2022data} proposed the constrained matrix
zonotope (CMZ) to tighten the MZ model set by enforcing
\emph{kernel-consistency} constraints induced by the data regressor.
This subsection reviews their construction, which yields a constrained
matrix zonotope that is a strict subset of the MZ model set; the CMZ
formulation itself is \emph{not} a contribution of the present work.

\paragraph{Noise matrix zonotope construction.}
Assume that the single-step disturbance belongs to a zonotope
$\mathcal{W} = \langle 0,\, \{g_w^{(j)}\}_{j=1}^{p_w} \rangle
\subset \mathbb{R}^{n_x}$.
The stacked disturbance matrix
$W_- = [w(0)\;\cdots\; w(T\!-\!1)] \in \mathbb{R}^{n_x \times T}$
belongs to a matrix zonotope $\mathcal{M}_w$ constructed by treating each
time step independently.  Specifically, for each noise generator index
$j \in \{1,\dots,p_w\}$ and each time index $t \in \{1,\dots,T\}$, the
rank-one generator matrix is defined as
\begin{equation}\label{eq:noise_gen_construction}
G_{w}^{(j,t)} := g_w^{(j)}\,e_t^\top \in \mathbb{R}^{n_x \times T},
\end{equation}
where $e_t$ denotes the $t$-th canonical basis vector in $\mathbb{R}^T$.
The full noise matrix zonotope is then
\begin{equation}\label{eq:Mw_full}
\mathcal{M}_w =
\Big\langle\, 0,\; \big\{G_{w}^{(j,t)}\big\}_{j=1,\,t=1}^{p_w,\,T}
\Big\rangle,
\end{equation}
with center $C_w = 0$ and $\kappa = p_w \cdot T$ generators.
Each coefficient $\beta_{(j,t)} \in [-1,1]$ scales one noise direction
$g_w^{(j)}$ at one time step $t$, so the disturbance at time $t$ is
$w(t) = \sum_{j=1}^{p_w} \beta_{(j,t)}\, g_w^{(j)}$, which correctly
ranges over $\mathcal{W}$ independently for each $t$.

The set $\mathcal{N} = X_+ - \mathcal{M}_w$ (cf.\
\eqref{eq:denoised_set}) then has center $C_n = X_+$ and generators
$G_\ell = -G_{w,\ell}$.

\paragraph{Kernel-consistency identity.}
Let $\Phi_\perp \in \mathbb{R}^{T \times r}$ be a basis for the right
nullspace of $\Phi$, i.e., $\Phi \Phi_\perp = 0$, where
$r := T - \mathrm{rank}(\Phi)$.
Note that $r = T - d > 0$ whenever $T > d = n_x + n_u$, which is a
necessary condition for any meaningful kernel-consistency constraint.
From~\eqref{eq:data_relation}, we have
\begin{equation}\label{eq:N_def}
N_{\mathrm{tr}} := X_+ - W_- = M_{\mathrm{tr}}\,\Phi.
\end{equation}
Multiplying both sides on the right by $\Phi_\perp$ yields the
kernel-consistency constraint
\begin{equation}\label{eq:kernel_constraint}
N_{\mathrm{tr}}\,\Phi_\perp = M_{\mathrm{tr}}\,\Phi\,\Phi_\perp = 0.
\end{equation}
Equation~\eqref{eq:kernel_constraint} states that the true data
matrix without noise lies in the left nullspace of $\Phi_\perp^\top$.
Consequently, the true data matrix without noise belongs not merely to
$\mathcal{N}$, but to the subset
\begin{equation}\label{eq:N0_def}
\mathcal{N}_0 := \{N \in \mathcal{N} \mid N\Phi_\perp = 0\}.
\end{equation}

\paragraph{CMZ representation of $\mathcal{N}_0$.}
Any $N \in \mathcal{N}$ can be written as
$N(\beta) = C_n + \sum_{\ell=1}^\kappa \beta_\ell G_\ell$ with
$\beta \in [-1,1]^\kappa$.
Imposing $N(\beta)\Phi_\perp = 0$ yields the matrix equation
\begin{equation}\label{eq:beta_constraints_matrix}
\sum_{\ell=1}^\kappa \beta_\ell\,(G_\ell\Phi_\perp)
= -C_n\Phi_\perp.
\end{equation}
This constitutes a system of $n_x \times r$ scalar equations in $\kappa$
unknowns.
To express~\eqref{eq:beta_constraints_matrix} as standard linear equations
in $\beta$, the $\mathrm{vec}$ operator is applied to both sides, giving
\begin{equation}\label{eq:vec_constraints}
A_{\mathrm{cmz}} :=
\begin{bmatrix}
\mathrm{vec}(G_1\Phi_\perp) & \cdots &
\mathrm{vec}(G_\kappa\Phi_\perp)
\end{bmatrix}
\in \mathbb{R}^{(n_x r) \times \kappa},
\end{equation}
and $b_{\mathrm{cmz}} := -\mathrm{vec}(C_n\Phi_\perp)
\in \mathbb{R}^{n_x r}$.
Then~\eqref{eq:beta_constraints_matrix} is equivalent to
$A_{\mathrm{cmz}}\,\beta = b_{\mathrm{cmz}}$.

\paragraph{Block-constraint form.}
Equivalently, the matrix-valued constraint blocks can be retained directly:
\begin{align}\label{eq:cmz_block_constraint}
&\textstyle\sum_{\ell=1}^{\kappa}\beta_\ell\, A_\ell^{\mathrm{blk}} = B^{\mathrm{blk}},\nonumber\\
&A_\ell^{\mathrm{blk}} := G_\ell\Phi_\perp \in \mathbb{R}^{n_x \times r},
\quad
B^{\mathrm{blk}} := -C_n\Phi_\perp.
\end{align}
This block form is the representation used in MATLAB via a cell array
$\{A_\ell^{\mathrm{blk}}\}_{\ell=1}^\kappa$ and right-hand side
$B^{\mathrm{blk}}$.

The kernel-consistent set admits the constrained matrix zonotope
description (Definition~\ref{def:conmatzono})
\begin{equation}\label{eq:cmz_model}
\mathcal{N}_0 = \bigg\{
C_n + \sum_{\ell=1}^\kappa \beta_\ell G_\ell \;\bigg|\;
\beta \in [-1,1]^\kappa,\;
A_{\mathrm{cmz}}\beta = b_{\mathrm{cmz}}\bigg\}.
\end{equation}

\paragraph{CMZ model set.}
Mapping $\mathcal{N}_0$ through a right inverse $H$ yields
\begin{multline}\label{eq:modelset_cmz}
\mathcal{M}_\Sigma^{\mathrm{CMZ}}(H) := \mathcal{N}_0\,H
= \bigg\{
\bigg(C_n + \textstyle\sum_{\ell=1}^\kappa \beta_\ell G_\ell\bigg)H
\;\bigg|\\
\beta \in [-1,1]^\kappa,\;
A_{\mathrm{cmz}}\beta = b_{\mathrm{cmz}}\bigg\}.
\end{multline}
The linear map $N \mapsto NH$ does not alter the coefficient vector $\beta$;
hence the constraints~\eqref{eq:vec_constraints} remain constraints on the
same $\beta$ after right multiplication by $H$, while the center and
generators become $C_n H$ and $G_\ell H$, respectively.
By construction,
$\mathcal{N}_0 \subseteq \mathcal{N}$, and therefore
\begin{equation}\label{eq:subset_relation}
\mathcal{M}_\Sigma^{\mathrm{CMZ}}(H) \subseteq
\mathcal{M}_\Sigma^{\mathrm{MZ}}(H).
\end{equation}

\begin{remark}[Constraint dimensions and effectiveness]\label{rem:constraint_dim}
The constraint matrix $A_{\mathrm{cmz}} \in \mathbb{R}^{(n_x r) \times \kappa}$
has $n_x r = n_x(T - d)$ rows and $\kappa = p_w T$ columns.
For the CMZ to be strictly tighter than the MZ, it is necessary that
$\mathrm{rank}(A_{\mathrm{cmz}}) > 0$, which holds whenever $r > 0$
(i.e., $T > d$) and the noise generators are not aligned with the
nullspace of $\Phi$.  In practice, the number of effective constraints
grows with $T - d$, so collecting more data than the minimum $T = d$
required for full row rank of $\Phi$ directly improves the tightening
provided by the CMZ.
\end{remark}

\subsection{Generator-Norm Proxy and Row-Norm Right Inverse}\label{sec:row_norm}

The size of a matrix zonotope
$\mathcal{M} = \langle C, \{G_\ell\}_{\ell=1}^\kappa \rangle$ is
quantified by the generator-norm proxy
\begin{equation}\label{eq:gen_proxy}
\mathsf{V}(\mathcal{M}) := \sum_{\ell=1}^\kappa \norm{G_\ell}_F.
\end{equation}

\paragraph{Stacked-noise structure.}
Assume that $\mathcal{W} = \langle 0, \{g_w^{(j)}\}_{j=1}^{p_w} \rangle$
is a zonotope for $w(k)$, and construct $\mathcal{M}_w$ by stacking
independent copies across time.  Each generator of $\mathcal{M}_w$ is then
the rank-one matrix
\begin{equation}\label{eq:noise_gen_rank1}
G_w^{(j,t)} = g_w^{(j)}\,e_t^\top \in \mathbb{R}^{n_x \times T},
\end{equation}
where $e_t$ is the $t$-th canonical basis vector.
Right-multiplying by $H$ yields
\begin{equation}\label{eq:noise_gen_push}
G_w^{(j,t)}H = g_w^{(j)}\,(e_t^\top H) = g_w^{(j)}\,h_t^\top,
\qquad h_t^\top := e_t^\top H.
\end{equation}
Using $\norm{ab^\top}_F = \norm{a}_2\norm{b}_2$ for rank-one matrices,
\begin{equation}\label{eq:frob_factor}
\norm{G_w^{(j,t)}H}_F = \norm{g_w^{(j)}}_2\,\norm{h_t}_2
= \norm{g_w^{(j)}}_2\,\norm{H_{t,:}}_2.
\end{equation}
The disturbance-induced portion of the proxy
of $\mathcal{M}_\Sigma$ in~\eqref{eq:modelset_mz} therefore factorizes as
\begin{equation}\label{eq:proxy_factor}
\sum_{j=1}^{p_w}\sum_{t=1}^{T}\norm{G_w^{(j,t)}H}_F
= \bigg(\sum_{j=1}^{p_w}\norm{g_w^{(j)}}_2\bigg)
  \bigg(\sum_{t=1}^{T}\norm{H_{t,:}}_2\bigg).
\end{equation}

\paragraph{Row-norm right inverse (SOCP)}
Because $\sum_j\norm{g_w^{(j)}}_2$ depends only on the noise bound,
minimizing~\eqref{eq:proxy_factor} reduces to
\begin{equation}\label{eq:socp}
H_{\mathrm{row}} \in
\arg\min_{H \in \mathbb{R}^{T \times d}}
\sum_{t=1}^T \norm{H_{t,:}}_2
\quad \text{s.t.}\quad \Phi H = I_d.
\end{equation}
Problem~\eqref{eq:socp} is a second-order cone program (SOCP) and can be
solved with standard convex optimization solvers.

For comparison, the pseudoinverse $H_{\mathrm{pinv}} = \Phi^\dagger$
is the minimum-Frobenius-norm right inverse.
The relationship between the two objectives is
\begin{equation}\label{eq:row_frob_bounds}
\norm{H}_F \le \sum_{t=1}^T \norm{H_{t,:}}_2
\le \sqrt{T}\,\norm{H}_F,
\end{equation}
which yields the bound
\begin{equation}\label{eq:gamma_bounds}
\norm{\Phi^\dagger}_F \le
\min_{\Phi H = I} \sum_t \norm{H_{t,:}}_2
\le \sqrt{T}\,\norm{\Phi^\dagger}_F.
\end{equation}
Thus, input design that improves the conditioning of $\Phi$ reduces both
objectives and consequently shrinks the model set.

\begin{lemma}[Proxy monotonicity under CMZ constraints]\label{lem:proxy_mono}
Let $\mathcal{M}^{\mathrm{MZ}} = \langle C, \{G_\ell\}_{\ell=1}^\kappa \rangle$
be a matrix zonotope and let
$\mathcal{M}^{\mathrm{CMZ}} = \langle C, \{G_\ell\}_{\ell=1}^\kappa,
A_{\mathrm{cmz}}, b_{\mathrm{cmz}} \rangle$
be the corresponding constrained matrix zonotope.  Then
$\mathcal{M}^{\mathrm{CMZ}} \subseteq \mathcal{M}^{\mathrm{MZ}}$.
Moreover, any set-valued function that is monotone with respect to set
inclusion preserves this ordering:  in particular,
$\mathcal{M}^{\mathrm{CMZ}} \mathcal{Z} \oplus \mathcal{W}
\subseteq \mathcal{M}^{\mathrm{MZ}} \mathcal{Z} \oplus \mathcal{W}$
for any zonotope $\mathcal{Z}$ and noise set $\mathcal{W}$.
\end{lemma}

\begin{compactproof}
The feasible set of $\beta$ for the CMZ is
$\mathcal{B}_c := \{\beta \in [-1,1]^\kappa \mid A_{\mathrm{cmz}}\beta
= b_{\mathrm{cmz}}\} \subseteq [-1,1]^\kappa =: \mathcal{B}$.
Because every matrix in $\mathcal{M}^{\mathrm{CMZ}}$ corresponds to some
$\beta \in \mathcal{B}_c \subseteq \mathcal{B}$, it is also contained in
$\mathcal{M}^{\mathrm{MZ}}$.  The second claim follows because the
set-valued map $\mathcal{M} \mapsto \mathcal{M}\mathcal{Z} \oplus \mathcal{W}$
is monotone with respect to set inclusion.
\end{compactproof}

\subsection{Online A-Optimal Input Design over Constrained
Zonotopes}\label{sec:input_design}

Inputs are designed online to improve the regressor matrix $\Phi$
in~\eqref{eq:Phi_def} without knowledge of
$(A_{\mathrm{tr}}, B_{\mathrm{tr}})$.
The regressor vector at time $k$ is defined as
\begin{equation}\label{eq:sk_def}
s_k := \begin{bmatrix}x(k)\\ u(k)\end{bmatrix} \in \mathbb{R}^d,
\qquad d := n_x + n_u,
\end{equation}
and the (regularized) information matrix is
\begin{equation}\label{eq:info_matrix}
S_k := \delta I_d + \sum_{t=0}^{k-1} s_t s_t^\top,
\qquad \delta > 0.
\end{equation}
The columns of $\Phi$ are exactly $\{s_t\}_{t=0}^{T-1}$, so
$\Phi\Phi^\top = \sum_{t=0}^{T-1} s_t s_t^\top = S_T - \delta I_d$.
Hence $S_T \succ 0$ and
$\sigma_{\min}(\Phi)^2 \ge \lambda_{\min}(S_T) - \delta$,
linking the information matrix to the conditioning of $\Phi$.

\paragraph{Greedy A-optimal criterion.}
The global design objective is to minimize $\mathrm{tr}(S_T^{-1})$
(A-optimality), which directly minimizes the model-set proxy
$\norm{\Phi^\dagger}_F^2 = \mathrm{tr}((\Phi\Phi^\top)^{-1})$.
Using the Sherman--Morrison rank-1 update formula for
$S_{k+1} = S_k + s_k s_k^\top$,
\begin{equation}\label{eq:sm_update}
S_{k+1}^{-1} = S_k^{-1}
- \frac{S_k^{-1} s_k s_k^\top S_k^{-1}}{1 + s_k^\top S_k^{-1} s_k}.
\end{equation}
Taking the trace of both sides yields
\begin{equation}\label{eq:trace_decrease}
\mathrm{tr}(S_{k+1}^{-1})
= \mathrm{tr}(S_k^{-1})
- \frac{s_k^\top S_k^{-2} s_k}{1 + s_k^\top S_k^{-1} s_k},
\end{equation}
where the identity
$\mathrm{tr}(S_k^{-1} s_k s_k^\top S_k^{-1}) = s_k^\top S_k^{-2} s_k$
has been used.  A one-step greedy A-optimal policy therefore
maximizes the decrease in $\mathrm{tr}(S_k^{-1})$:
\begin{align}\label{eq:aopt}
u(k) &\in \arg\max_{u \in \mathcal{U}}\;
\Delta_A(u),\nonumber\\
\Delta_A(u) &:=
\frac{\begin{bmatrix}x(k)\\ u\end{bmatrix}^{\!\top}
S_k^{-2}
\begin{bmatrix}x(k)\\ u\end{bmatrix}}
{1 + \begin{bmatrix}x(k)\\ u\end{bmatrix}^{\!\top}
S_k^{-1}
\begin{bmatrix}x(k)\\ u\end{bmatrix}},\nonumber\\
S_{k+1} &= S_k + s_k s_k^\top.
\end{align}
The numerator $s^\top S_k^{-2} s$ strongly penalizes directions along which
the information matrix has small eigenvalues, while the denominator
$1 + s^\top S_k^{-1} s$ provides normalization arising from the rank-1
update algebra.

\paragraph{Optimization over a constrained zonotope.}
Since $\mathcal{U} = \langle c_u, G_u, A_u, b_u \rangle$ is
parameterized by factors $\xi$, substituting $u = c_u + G_u \xi$ into
\eqref{eq:aopt} yields a fractional quadratic program:
\begin{equation}\label{eq:aopt_factor}
\begin{aligned}
\max_{\xi \in \mathbb{R}^m}\quad &
\frac{\xi^\top Q_2\, \xi + 2\,q_2^\top \xi + c_2}
{1 + \xi^\top Q_1\, \xi + 2\,q_1^\top \xi + c_1}\\
\mathrm{s.t.}\quad &
\norm{\xi}_\infty \le 1,\qquad A_u\xi = b_u,
\end{aligned}
\end{equation}
where $Q_j, q_j, c_j$ ($j=1,2$) are obtained by partitioning $S_k^{-1}$
and $S_k^{-2}$ conformally with the $(x,u)$ block structure.
\begin{enumerate}
\item \emph{Global exploration:}  $N_{\mathrm{cand}}$ feasible candidates
  are sampled uniformly from $\mathcal{U}$ and evaluated.
\item \emph{Local refinement:}  starting from the best candidate, SQP is
  run in the $\xi$-space subject to
  $\norm{\xi}_\infty \le 1$ and $A_u \xi = b_u$.
\end{enumerate}

\subsection{Reachable-Set Propagation}\label{sec:propagation}

Let $\widehat{\mathcal{R}}_0 = \mathcal{X}_0$ and define the lifted set
$\mathcal{Z}_k := \widehat{\mathcal{R}}_k \times \mathcal{U}_k$,
where $\mathcal{U}_k$ is the (possibly time-varying) input set used during
propagation.  For any model set $\mathcal{M}_\Sigma$ (MZ or CMZ), the
one-step reachable-set over-approximation is
\begin{equation}\label{eq:reach_update}
\widehat{\mathcal{R}}_{k+1}
= \mathcal{M}_\Sigma\,\mathcal{Z}_k \oplus \mathcal{W}.
\end{equation}

\paragraph{Matrix zonotope--zonotope multiplication.}
When $\mathcal{M}_\Sigma = \langle C, \{G_\ell\}_{\ell=1}^\kappa \rangle$
is a standard MZ and
$\mathcal{Z}_k = \langle c_z, G_z \rangle$ is a zonotope, the product
$\mathcal{M}_\Sigma\, \mathcal{Z}_k$ is
over-approximated by~\cite{amr23reachable}
\begin{multline}\label{eq:mz_zono_mult}
\mathcal{M}_\Sigma\, \mathcal{Z}_k
\subseteq \Big\langle C\,c_z,\;
\{C\, g_{z,i}\}_{i=1}^{n_z}
\cup
\{G_\ell\, c_z\}_{\ell=1}^\kappa \\
\cup\;
\{G_\ell\, g_{z,i}\}_{\ell,i}
\Big\rangle,
\end{multline}
where $g_{z,i}$ denotes the columns of $G_z$ and $n_z$ is the number of
generators of $\mathcal{Z}_k$.
The inclusion (rather than equality) arises because the bilinear products
$\beta_\ell \alpha_i$ of the MZ and zonotope coefficients are treated as
independent factors in $[-1,1]$, which enlarges the resulting set.
This is, however, a {\em sound} outer approximation that preserves the
reachable-set containment guarantee of Lemma~\ref{lem:soundness}.
The total
number of resulting generators is $n_z + \kappa + \kappa n_z$, which
grows quadratically in $\kappa$ and $n_z$, necessitating zonotope order
reduction~\cite{kopetzki2017methods} after each propagation step.

When $\mathcal{M}_\Sigma$ is a CMZ, the product
$\mathcal{M}_\Sigma\,\mathcal{Z}_k$ is over-approximated by a
constrained zonotope via Proposition~\ref{prop:cmz_product}, which
preserves the linear constraints on the CMZ coefficients $\beta$.

\subsection{Main Result}\label{sec:main_result}

The following theorem consolidates the key contributions of this paper:
right-inverse optimization and A-optimal input design jointly reduce
the conservatism of data-driven reachable-set over-approximations while
preserving soundness.

\begin{theorem}[Tighter Over-Approximation via Input Design and
Right-Inverse Optimization]\label{thm:main}
Consider system~\eqref{eq:lti_true} with bounded noise
$w(k) \in \mathcal{W}$.
Let $\Phi^{\mathrm{A}}$ and $\Phi^{\mathrm{R}}$ be regressor matrices
obtained from A-optimal designed and random inputs, respectively, both
with full row rank.
Let $H_{\mathrm{row}}$ denote the SOCP row-norm-minimizing right
inverse~\eqref{eq:socp} and $H_{\mathrm{pinv}} := \Phi^\dagger$ the
pseudoinverse.
Then the following hold.
\begin{enumerate}
\item[\emph{(i)}] \emph{(Soundness.)}
  $M_{\mathrm{tr}} \in \mathcal{M}_\Sigma(H)$ for any right inverse $H$
  satisfying $\Phi H = I_d$, and consequently
  $\mathcal{R}_k \subseteq \widehat{\mathcal{R}}_k$ for all $k \ge 0$.
\item[\emph{(ii)}] \emph{(Right-inverse tightening.)}
  For a fixed regressor $\Phi$, the generator-norm
  proxy~\eqref{eq:gen_proxy} satisfies
  \begin{equation}\label{eq:main_ri}
  \mathsf{V}\!\big(\mathcal{M}_\Sigma(H_{\mathrm{row}})\big)
  \le
  \mathsf{V}\!\big(\mathcal{M}_\Sigma(H_{\mathrm{pinv}})\big).
  \end{equation}
\item[\emph{(iii)}] \emph{(Input-design tightening.)}
  A-optimal designed inputs reduce the pseudoinverse norm:
  $\norm{(\Phi^{\mathrm{A}})^\dagger}_F
  \le \norm{(\Phi^{\mathrm{R}})^\dagger}_F$,
  which in turn reduces the generator-norm proxy of the model set
  for any choice of right inverse.
\item[\emph{(iv)}] \emph{(Combined effect.)}
  The two improvements are complementary.
  Combining A-optimal inputs with the SOCP right inverse yields
  \begin{equation}\label{eq:main_combined}
  \mathsf{V}\!\big(\mathcal{M}_\Sigma^{\mathrm{A}}(H_{\mathrm{row}})\big)
  \le
  \mathsf{V}\!\big(\mathcal{M}_\Sigma^{\mathrm{A}}(H_{\mathrm{pinv}})\big)
  \le
  \mathsf{V}\!\big(\mathcal{M}_\Sigma^{\mathrm{R}}(H_{\mathrm{pinv}})\big),
  \end{equation}
  where superscripts $\mathrm{A}$ and $\mathrm{R}$ denote designed and
  random inputs, respectively.  Because the reachable-set propagation
  operator~\eqref{eq:reach_update} is monotone with respect to the
  model-set size, the resulting over-approximation
  $\widehat{\mathcal{R}}_k^{\mathrm{A}}(H_{\mathrm{row}})$ is the
  tightest among all four combinations.
\end{enumerate}
\end{theorem}

\begin{compactproof}
Part~(i) follows from Lemma~\ref{lem:soundness}.
Part~(ii): by~\eqref{eq:proxy_factor}, the disturbance-induced proxy is
proportional to $\sum_t \norm{H_{t,:}}_2$, which
$H_{\mathrm{row}}$ minimizes by construction~\eqref{eq:socp};
hence $\mathsf{V}(\mathcal{M}_\Sigma(H_{\mathrm{row}}))
\le \mathsf{V}(\mathcal{M}_\Sigma(H))$ for any right inverse $H$,
including $H_{\mathrm{pinv}}$.
Part~(iii): the A-optimal criterion minimizes
$\mathrm{tr}((\Phi\Phi^\top)^{-1}) = \norm{\Phi^\dagger}_F^2$;
by the sandwich bound~\eqref{eq:row_norm_sandwich}, a smaller
$\norm{\Phi^\dagger}_F$ reduces the row-norm sum for any right inverse.
Part~(iv): the first inequality in~\eqref{eq:main_combined} is
part~(ii) applied to $\Phi^{\mathrm{A}}$; the second is
part~(iii) applied to $H_{\mathrm{pinv}}$.
Monotonicity of the propagation operator then yields
$\widehat{\mathcal{R}}_k^{\mathrm{A}}(H_{\mathrm{row}})
\subseteq \widehat{\mathcal{R}}_k^{\mathrm{R}}(H_{\mathrm{pinv}})$
for all $k$.
\end{compactproof}

\subsection{Extension to Piecewise Affine Systems}\label{sec:pwa}

Consider a piecewise affine (PWA) system with $Q$ modes~\cite{xie2025data}:
\begin{multline}\label{eq:pwa_system}
x(k{+}1) = A_q\,x(k) + B_q\,u(k) + w(k),\\
x(k) \in \mathcal{P}_q,
\quad q = 1,\dots,Q,
\end{multline}
where $\{\mathcal{P}_q\}_{q=1}^Q$ is a polyhedral partition of the state
space.  The mode-specific system matrices $[A_q\;\; B_q]$ are unknown;
only the partition geometry is assumed to be known.

\paragraph{Per-mode data partitioning.}
For each mode $q$, all data transitions satisfying $x(k) \in \mathcal{P}_q$
are collected into separate data matrices
$(X_{-,q},\, U_{-,q},\, X_{+,q})$.
The mode-specific regressor is
$\Phi_q = \big[\begin{smallmatrix} X_{-,q} \\ U_{-,q}\end{smallmatrix}\big]
\in \mathbb{R}^{d \times T_q}$,
where $T_q$ is the number of transitions in mode $q$.
A separate constrained matrix zonotope
$\mathcal{M}_{\Sigma,q}^{\mathrm{CMZ}}$ is then constructed for each
mode using the procedure described in Section~\ref{sec:cmz}.

\paragraph{Guard splitting.}
At each propagation step, the current reachable set
$\widehat{\mathcal{R}}_k$ may overlap multiple mode regions.
For each guard surface $\mathcal{H}_q := \partial \mathcal{P}_q$
(e.g., a hyperplane $h^\top x = c$), the set
$\widehat{\mathcal{R}}_k$ is split into fragments:
\begin{equation}\label{eq:guard_split}
\widehat{\mathcal{R}}_k^{(q)}
:= \widehat{\mathcal{R}}_k \cap \mathcal{P}_q,
\qquad q = 1,\dots,Q.
\end{equation}
When $\widehat{\mathcal{R}}_k$ is a zonotope (or constrained zonotope)
and $\mathcal{P}_q$ is a halfspace, the intersection
$\widehat{\mathcal{R}}_k^{(q)}$ is a constrained
zonotope~\cite{scott2016constrained}.  Each fragment is then propagated
under its respective mode:
\begin{equation}\label{eq:pwa_propagation}
\widehat{\mathcal{R}}_{k+1}^{(q)}
= \mathcal{M}_{\Sigma,q}\,
\big(\widehat{\mathcal{R}}_k^{(q)} \times \mathcal{U}\big)
\oplus \mathcal{W},
\end{equation}
and the full reachable set at step $k+1$ is
$\widehat{\mathcal{R}}_{k+1} = \bigcup_{q=1}^Q
\widehat{\mathcal{R}}_{k+1}^{(q)}$.
For $Q=2$ modes, this produces a binary tree with up to $2^k$ branches
at step $k$, although branches for which
$\widehat{\mathcal{R}}_k^{(q)} = \emptyset$ are pruned.

The model-based reference uses
a mixed logical dynamical (MLD) formulation with hybrid zonotope
propagation~\cite{bemporad1999control,bird2023hybrid}.

%% file: sections/theory.tex
\section{Theoretical Results}\label{sec:theory}

This section establishes the soundness guarantees inherited from
prior work and presents the main theoretical contributions of this
paper: the row-norm right-inverse bounds and the A-optimal
input-design proxy reduction.

\begin{lemma}[Data-Driven Reachable-Set Soundness~\cite{amr23reachable,alanwar2022data}]\label{lem:soundness}
Suppose $w(k) \in \mathcal{W}$ for all $k$ and $\Phi$ has full row rank.
Let $H$ satisfy $\Phi H = I_d$.  Then:
\begin{enumerate}
\item[\emph{(i)}]
  $M_{\mathrm{tr}} \in
  \mathcal{M}_\Sigma^{\mathrm{CMZ}}(H) \subseteq
  \mathcal{M}_\Sigma^{\mathrm{MZ}}(H)$;
\item[\emph{(ii)}]
  $\mathcal{R}_k \subseteq
  \widehat{\mathcal{R}}_k^{\mathrm{CMZ}}(H) \subseteq
  \widehat{\mathcal{R}}_k^{\mathrm{MZ}}(H)$
  for all $k \ge 0$.
\end{enumerate}
\end{lemma}

\begin{compactproof}
(i)~From~\eqref{eq:data_relation}, $N_{\mathrm{tr}} = X_+ - W_-
= M_{\mathrm{tr}}\Phi$.
Because $W_- \in \mathcal{M}_w$, there exists
$\beta^\star \in [-1,1]^\kappa$ such that
$N_{\mathrm{tr}} = C_n + \sum_\ell \beta_\ell^\star G_\ell \in \mathcal{N}$.
The kernel-consistency identity $N_{\mathrm{tr}}\Phi_\perp = 0$
yields $A_{\mathrm{cmz}}\beta^\star = b_{\mathrm{cmz}}$, so
$N_{\mathrm{tr}} \in \mathcal{N}_0 \subseteq \mathcal{N}$.
Right-multiplying by $H$ gives
$M_{\mathrm{tr}} = N_{\mathrm{tr}} H
\in \mathcal{N}_0 H = \mathcal{M}_\Sigma^{\mathrm{CMZ}}(H)
\subseteq \mathcal{N} H = \mathcal{M}_\Sigma^{\mathrm{MZ}}(H)$.

(ii)~Induction on $k$: the base case
$\mathcal{R}_0 = \mathcal{X}_0 = \widehat{\mathcal{R}}_0$ is immediate.
For the inductive step, $M_{\mathrm{tr}} \in \mathcal{M}_\Sigma$ by~(i),
so $x(k{+}1) = M_{\mathrm{tr}} z + w(k)
\in \mathcal{M}_\Sigma\mathcal{Z}_k \oplus \mathcal{W}
= \widehat{\mathcal{R}}_{k+1}$.
The CMZ $\subseteq$ MZ ordering follows from
Lemma~\ref{lem:proxy_mono}.
\end{compactproof}

\begin{theorem}[Row-Norm Bounds]\label{prop:row_norm}
For any full-row-rank $\Phi \in \mathbb{R}^{d \times T}$, let
$\gamma^\star := \min_{\Phi H = I}\sum_{t}\norm{H_{t,:}}_2$.  Then
\begin{equation}\label{eq:row_norm_sandwich}
\norm{\Phi^\dagger}_F \le \gamma^\star
\le \sqrt{T}\,\norm{\Phi^\dagger}_F.
\end{equation}
\end{theorem}

\begin{compactproof}
\emph{Left bound:}
The pseudoinverse $\Phi^\dagger$ satisfies $\Phi\Phi^\dagger = I_d$ and
is therefore a feasible right inverse.  For any matrix $A$,
$\norm{A}_F = (\sum_t \norm{A_{t,:}}_2^2)^{1/2}
\le \sum_t \norm{A_{t,:}}_2$
by the norm comparison $\ell_2 \le \ell_1$ applied to the vector
$(\norm{A_{1,:}}_2, \dots, \norm{A_{T,:}}_2)$.
Because $\Phi^\dagger$ minimizes $\norm{H}_F$ among all right inverses $H$,
and for any $H$,
$\norm{H}_F \le \sum_t \norm{H_{t,:}}_2$
(by the $\ell_2 \le \ell_1$ norm comparison),
it follows that $\norm{\Phi^\dagger}_F \le \norm{H^\star}_F
\le \sum_t \norm{H^\star_{t,:}}_2 = \gamma^\star$,
where $H^\star$ denotes the row-norm-optimal right inverse.

\emph{Right bound:}
By the Cauchy--Schwarz inequality in $\mathbb{R}^T$,
$\sum_t \norm{H_{t,:}}_2 \le \sqrt{T}\,(\sum_t \norm{H_{t,:}}_2^2)^{1/2}
= \sqrt{T}\,\norm{H}_F$.
Evaluating at $H^\star$ (the row-norm minimizer) and using
$\norm{H^\star}_F \ge \norm{\Phi^\dagger}_F$ gives
$\gamma^\star \le \sqrt{T}\,\norm{H^\star}_F$.
In addition, $\norm{H^\star}_F \ge \norm{\Phi^\dagger}_F$, so the bound
is obtained by noting that for $H = \Phi^\dagger$,
$\gamma^\star \le \sum_t \norm{(\Phi^\dagger)_{t,:}}_2
\le \sqrt{T}\,\norm{\Phi^\dagger}_F$.
\end{compactproof}

\begin{theorem}[A-Optimal Design Reduces Proxy]%
\label{prop:aopt_proxy}
Let $\Phi^{\mathrm{R}}$ and $\Phi^{\mathrm{A}}$ be regressor matrices
obtained from random and A-optimal designed inputs, respectively, both
with full row rank.  If
$\mathrm{tr}\!\big((\Phi^{\mathrm{A}}(\Phi^{\mathrm{A}})^\top)^{-1}\big)
\le
\mathrm{tr}\!\big((\Phi^{\mathrm{R}}(\Phi^{\mathrm{R}})^\top)^{-1}\big)$,
then
\begin{equation}\label{eq:aopt_pinv_bound}
\norm{(\Phi^{\mathrm{A}})^\dagger}_F
\le \norm{(\Phi^{\mathrm{R}})^\dagger}_F.
\end{equation}
\end{theorem}

\begin{compactproof}
For any full-row-rank $\Phi \in \mathbb{R}^{d \times T}$,
$\norm{\Phi^\dagger}_F^2
= \mathrm{tr}(\Phi^\top(\Phi\Phi^\top)^{-2}\Phi)
= \mathrm{tr}((\Phi\Phi^\top)^{-1})$.
The A-optimal design directly minimizes
$\mathrm{tr}((\Phi\Phi^\top)^{-1})$,
so $\norm{(\Phi^{\mathrm{A}})^\dagger}_F^2
\le \norm{(\Phi^{\mathrm{R}})^\dagger}_F^2$.
\end{compactproof}

\begin{corollary}[Ordering of Over-Approximations]\label{cor:ordering}
Under the assumptions of Lemma~\ref{lem:soundness},
for any right inverse $H$ with $\Phi H = I_d$:
\begin{equation}\label{eq:reach_ordering}
\mathcal{R}_k
\subseteq \widehat{\mathcal{R}}_k^{\mathrm{CMZ}}(H)
\subseteq \widehat{\mathcal{R}}_k^{\mathrm{MZ}}(H)
\quad \forall\, k \ge 0.
\end{equation}
Moreover, for a fixed model-set type and right inverse, replacing random
inputs with A-optimal designed inputs yields tighter
over-approximations through the proxy reduction established in
Theorem~\ref{prop:aopt_proxy}.
\end{corollary}

\begin{compactproof}
The inclusions follow directly from
Lemma~\ref{lem:soundness}\,(ii).
The input-design claim follows because A-optimal inputs reduce
$\norm{\Phi^\dagger}_F$ (Theorem~\ref{prop:aopt_proxy}), which
reduces the generator-norm proxy via~\eqref{eq:proxy_factor}, and
the propagation operator is monotone with respect to model-set
inclusion (Lemma~\ref{lem:proxy_mono}).
\end{compactproof}

\begin{figure*}[t]
\centering
\includegraphics[width=\textwidth]{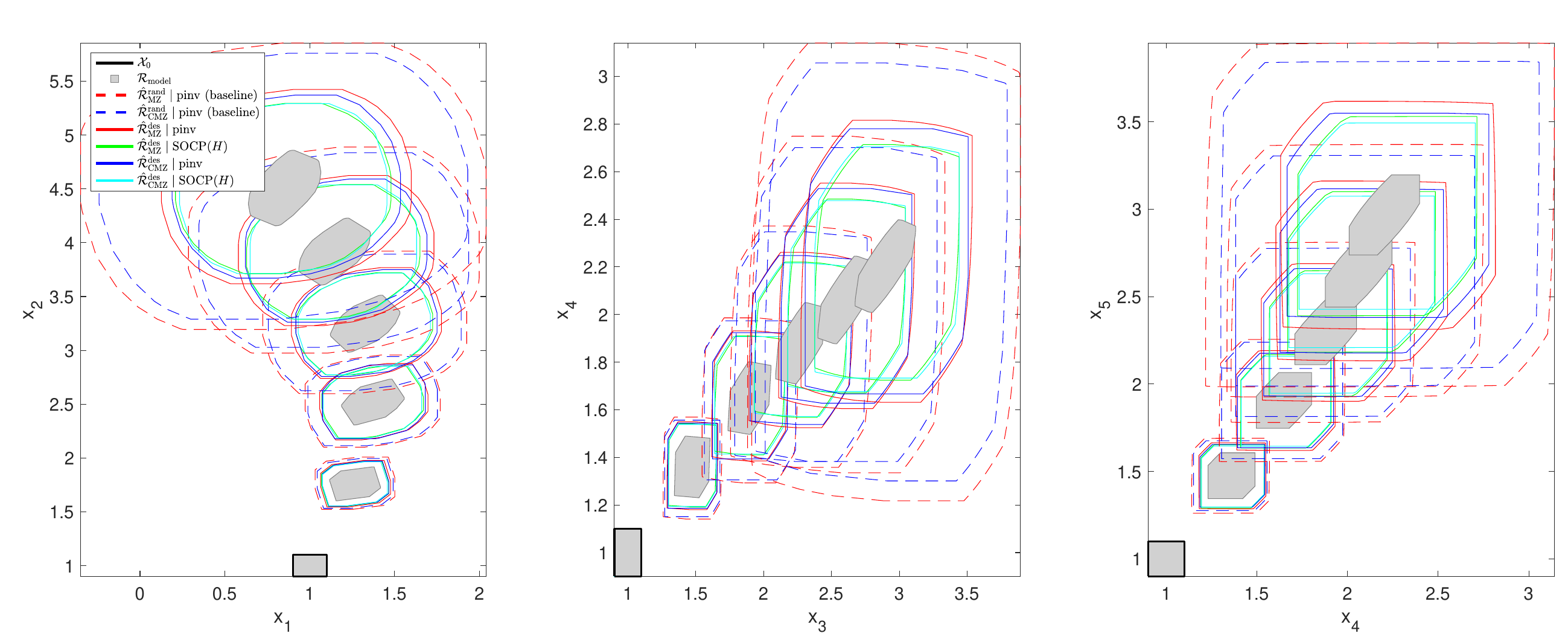}
\caption{Reachable-set comparison on the five-dimensional LTI system,
projected onto $(x_1,x_2)$, $(x_3,x_4)$, and $(x_4,x_5)$.
$\mathcal{R}_{\mathrm{model}}$: model-based ground truth (light gray filled).
Dashed lines show the random-input baselines from~\cite{alanwar2022data}:
($\hat{\mathcal{R}}_{\mathrm{MZ}}^{\mathrm{rand}}\!\mid\!\mathrm{pinv}$):
MZ with pseudoinverse (red dashed);
($\hat{\mathcal{R}}_{\mathrm{CMZ}}^{\mathrm{rand}}\!\mid\!\mathrm{pinv}$):
CMZ with pseudoinverse (blue dashed).
Solid lines show the proposed designed-input variants:
($\hat{\mathcal{R}}_{\mathrm{MZ}}^{\mathrm{des}}\!\mid\!\mathrm{pinv}$):
MZ with pseudoinverse (red solid);
($\hat{\mathcal{R}}_{\mathrm{MZ}}^{\mathrm{des}}\!\mid\!\mathrm{SOCP}(H)$):
MZ with SOCP right inverse (green);
($\hat{\mathcal{R}}_{\mathrm{CMZ}}^{\mathrm{des}}\!\mid\!\mathrm{pinv}$):
CMZ with pseudoinverse (blue solid);
($\hat{\mathcal{R}}_{\mathrm{CMZ}}^{\mathrm{des}}\!\mid\!\mathrm{SOCP}(H)$):
CMZ with SOCP right inverse (cyan).
All designed-input variants are visibly tighter than the corresponding
random-input baselines, and the combination
of CMZ with SOCP($H$) yields the tightest bound overall.}
\label{fig:lti_designed}
\end{figure*}

%% file: sections/experiments.tex
\section{Numerical Experiments}\label{sec:experiments}

All experiments are implemented in MATLAB R2024b with the Control System
Toolbox, the Optimization Toolbox, and CORA~2025~\cite{althoff2015introduction}.
Gurobi~13.0 serves as the LP solver for the PWA experiments.

\subsection{LTI System}\label{sec:lti_exp}

The five-dimensional continuous-time plant has state matrix
\begin{equation*}
A_c = \mathrm{diag}\!\left(
\begin{bmatrix}-1 &-4\\4 &-1\end{bmatrix},\;
\begin{bmatrix}-3 &1\\-1 &-3\end{bmatrix},\;
-2\right)
\end{equation*}
and input matrix $B_c = \mathbf{1}_{5 \times 3}$, discretized at
$\Delta t = 0.05$\,s.  The initial set $\mathcal{X}_0$ is a zonotope
centered at $\mathbf{1}_{5\times 1}$ with generator matrix $0.1 I_5$,
and the process noise $\mathcal{W}$ is a zero-centered zonotope with
generator matrix $0.005 I_5$.

The input set is the zonotope
$\mathcal{U} = \langle c_u, G_u \rangle$ with
$c_u = 10 \cdot \mathbf{1}_{3 \times 1}$,
\begin{equation*}
G_u = 10\begin{bmatrix}
6&1&1\\-2&7&-2\\0&1&-6
\end{bmatrix}.
\end{equation*}
Data are collected from $K = 12$ trajectories of $T_i = 5$ steps each
($T = 60$).

Reachable sets are propagated over 6 steps from $\mathcal{X}_0$ with
$\mathcal{U}_{\mathrm{prop}} = \{[10,\,5,\,-3]^\top\} \oplus
\mathrm{diag}(0.25, 0.15, 0.35)\,\mathcal{B}_\infty^3$.
Zonotope order reduction follows the Girard method with a maximum
order of 50 generators.
Four combinations of input quality (random versus designed) and
right-inverse selection (pseudoinverse versus SOCP row-norm minimizer)
are compared, each applied with both the MZ and
CMZ~\cite{alanwar2022data} model sets.
Fig.~\ref{fig:lti_designed} presents the results obtained with designed inputs;
dashed lines overlay the random-input baselines
from~\cite{alanwar2022data,amr23reachable} for comparison.

All data-driven reachable sets contain the model-based reachable set
computed from the true $(A_{\mathrm{tr}}, B_{\mathrm{tr}})$, confirming
soundness (Lemma~\ref{lem:soundness}).
Two consistent trends are visible in Fig.~\ref{fig:lti_designed}:
(i)~\emph{The SOCP right inverse tightens the model set:}
the SOCP variants (green, cyan) are contained
within the pseudoinverse variants (red, blue), validating
Theorem~\ref{prop:row_norm}.
(ii)~\emph{CMZ constraints further tighten the model set:}
CMZ-based sets (blue, cyan) are contained in their MZ counterparts
(red, green), as predicted by~\eqref{eq:subset_relation}.
All designed-input variants (solid) are uniformly tighter than the
random-input baselines (dashed), validating the input-design criterion.
The tightest over-approximation overall is
$\hat{\mathcal{R}}_{\mathrm{CMZ}}^{\mathrm{des}}\!\mid\!\mathrm{SOCP}(H)$
(cyan).

\paragraph{Quantitative comparison via volume.}
To complement the visual comparison, Table~\ref{tab:volume} reports the
volume of the reachable-set over-approximations at the final propagation
step for the MZ-based methods.
The volume of each zonotope is computed via the combinatorial
determinant formula, which sums the absolute values of the determinants
of all square submatrices of the generator
matrix~\cite{gover2010determinants}.
Because computing the exact volume of a constrained zonotope requires
vertex enumeration whose cost grows combinatorially with the number of
generators and constraints, only the unconstrained matrix-zonotope
variants are included.
The ratio column normalizes each volume by the model-based ground truth.

\begin{table}[t]
\centering
\caption{Volume of the reachable-set over-approximation at the final
propagation step (MZ methods only).  The ratio is relative to the
model-based reachable set.}
\label{tab:volume}
\begin{tabular}{lcc}
\hline
\textbf{Method} & \textbf{Volume} & \textbf{Ratio to Model} \\
\hline
Model                                                              & $1.62 \times 10^{-3}$ & $1.0\times$  \\
$\hat{\mathcal{R}}_{\mathrm{MZ}}^{\mathrm{rand}}\!\mid\!\mathrm{pinv}$ (baseline) & $1.04 \times 10^{-1}$ & $64.0\times$ \\
$\hat{\mathcal{R}}_{\mathrm{MZ}}^{\mathrm{des}}\!\mid\!\mathrm{pinv}$             & $6.83 \times 10^{-2}$ & $42.1\times$ \\
$\hat{\mathcal{R}}_{\mathrm{MZ}}^{\mathrm{des}}\!\mid\!\mathrm{SOCP}(H)$         & $2.77 \times 10^{-2}$ & $17.1\times$ \\
\hline
\end{tabular}
\end{table}

The random-input baseline produces a reachable set whose volume is
approximately $64\times$ that of the model-based ground truth.
Switching to A-optimal designed inputs while retaining the pseudoinverse
reduces this ratio to $42\times$, a $34\%$ reduction attributable
solely to improved data quality.
Replacing the pseudoinverse with the SOCP right inverse further
decreases the ratio to $17\times$, yielding a combined
$73\%$ volume reduction relative to the baseline.
These results confirm that the two proposed improvements---input design
and right-inverse optimization---provide substantial and complementary
reductions in conservatism.

\subsection{PWA System: Three-Method Comparison}\label{sec:pwa_exp}

The two-mode PWA system has modes
\begin{equation*}
A_1 = \begin{bmatrix}0.75&0.25\\-0.25&0.75\end{bmatrix},\quad
B_1 = \begin{bmatrix}-0.25\\-0.25\end{bmatrix}
\quad\text{for } x_1 \ge 0,
\end{equation*}
\begin{equation*}
A_2 = \begin{bmatrix}0.75&-0.25\\0.25&0.75\end{bmatrix},\quad
B_2 = \begin{bmatrix}0.25\\-0.25\end{bmatrix}
\quad\text{for } x_1 < 0.
\end{equation*}
The guard surface is the hyperplane $x_1 = 0$.
The initial set $\mathcal{X}_0$ is chosen such that trajectories cross the
guard within the 10-step propagation horizon.

Three methods are compared:
(i) model-based PWA propagation via hybrid zonotopes
($\mathcal{R}_{\mathrm{PWA}}$),
(ii) data-driven reachability with random inputs
($\hat{\mathcal{R}}_{\mathrm{rand}}\!\mid\!\mathrm{pinv}$), and
(iii) data-driven reachability with A-optimal designed inputs
($\hat{\mathcal{R}}_{\mathrm{des}}\!\mid\!\mathrm{SOCP}(H)$).
In both data-driven cases, constrained matrix zonotopes are
constructed for each mode and propagated using
Proposition~\ref{prop:cmz_product}.
The input zonotope is
$\mathcal{U} = \{-4\} \oplus 0.025\,\mathcal{B}_\infty^1$ and the noise
set is $\mathcal{W} = 0.005 I_2 \mathcal{B}_\infty^2$.

\begin{figure}[t]
\centering
\includegraphics[width=\columnwidth]{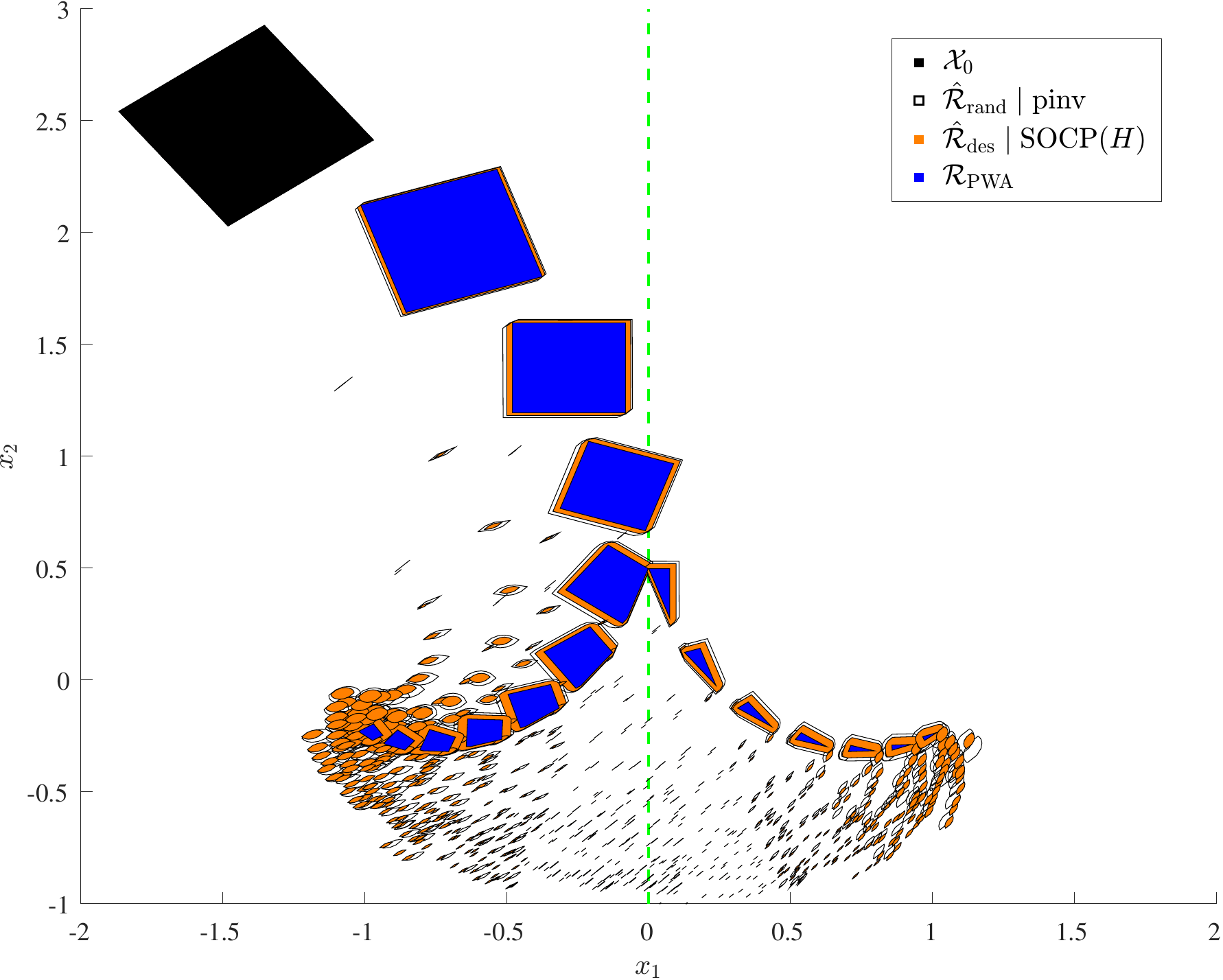}
\caption{PWA system reachable-set comparison over 10 steps.
Black filled: initial set $\mathcal{X}_0$.
$\hat{\mathcal{R}}_{\mathrm{rand}}\!\mid\!\mathrm{pinv}$:
random-input data-driven over-approximation (unfilled contours).
$\hat{\mathcal{R}}_{\mathrm{des}}\!\mid\!\mathrm{SOCP}(H)$:
designed-input data-driven over-approximation (orange filled).
$\mathcal{R}_{\mathrm{PWA}}$: model-based PWA reachable set (blue filled).
The guard surface $x_1 = 0$ is shown as a dashed line.
Both data-driven over-approximations soundly contain the model-based set,
and the designed-input variant produces a visibly tighter bound.}
\label{fig:pwa}
\end{figure}

Both data-driven methods produce over-approximations that contain the
model-based PWA reachable set $\mathcal{R}_{\mathrm{PWA}}$
(Fig.~\ref{fig:pwa}), confirming soundness.  The A-optimal variant
$\hat{\mathcal{R}}_{\mathrm{des}}$ yields a visibly tighter
over-approximation than
$\hat{\mathcal{R}}_{\mathrm{rand}}$.
Timing measurements indicate that matrix-zonotope propagation is the
fastest approach, while the MLD computation with hybrid zonotopes is the
most expensive due to the combinatorial nature of the mixed-integer
constraints.

%% file: sections/conclusion.tex
\section{Conclusion}\label{sec:conclusion}

This paper proposed a data-driven reachability analysis framework for discrete-time linear systems with unknown dynamics, combining constrained matrix zonotopes with right-inverse optimization and active input design. The results demonstrate that optimizing the right inverse and improving the informativeness of the collected data significantly reduce conservatism in reachable-set over-approximations. The approach was further applied to piecewise affine systems via mode-dependent data partitioning and hybrid zonotope propagation.
Future work will focus on addressing the coupling between piecewise affine regions and submodel parameters, extending input design to multi-step horizons, and further tightening the matrix-zonotope--zonotope product.